\documentclass{article}
\usepackage{spconf,amsmath,graphicx,hyperref}

\title{Neural Representations, Natural Connections: What Transfers From Human Speech Foundation Models to Animal Vocalizations?}
\name{\shortstack{
Tom{\'a}s Arias-Vergara$^{1}$\thanks{Corresponding author: tomas.arias@fau.de.}, Christopher Hauer$^{1,2}$, H{\'e}lo{\"i}se Brotier$^{3}$,
Elmar N{\"o}th$^{1}$, \\Andreas Maier$^{1}$, Lee Koren$^{3}$
}}
\address{$^{1}$Pattern Recognition Lab, Friedrich-Alexander-Universit{\"a}t Erlangen-N{\"u}rnberg, Germany\\
        $^{2}$Faculty of Electrical Engineering, Media and Computer Science, OTH Amberg-Weiden, Germany\\
        $^{3}$Faculty of Life Science, The Gonda Multidisciplinary Brain Research Center, Bar Ilan
        \\ University, Israel
        }

\begin{document}
%
\maketitle
\begin{abstract}
Speech-pretrained models have shown promise in animal bioacoustics, 
but the factors governing their cross-species transfer remain poorly understood.
We evaluate 15 frozen encoders spanning monolingual and multilingual speech,
speaker verification, animal bioacoustics, and general audio on caller
identification across four species and call-type classification across three.
Differences between the strongest speech- and animal-pretrained representations
range from $-0.027$ to $+0.119$ UAR; speech is significantly better in three of
seven settings ($p<0.01$) and not significantly different in the remaining four.
Controlled comparisons show no systematic advantage from increased language coverage or
animal-domain pretraining, while frozen speaker-verification embeddings
transfer poorly. The results also show that transfer is strongly layer-dependent: 
raw-waveform models peak early, whereas patch-spectrogram models peak deeper.
\end{abstract}
\section{Introduction}
\label{sec:intro}
Speech foundation models learn representations that transfer well beyond
the objectives used for pretraining. Self-supervised (SSL) models such as wav2vec 2.0~\cite{baevski2020wav2vec}, HuBERT~\cite{hsu2021hubert} and WavLM~\cite{chen2022wavlm}
encode not only linguistic content but also speaker and acoustic information, with different properties emerging across layers~\cite{pasad2021layerwise}.
This suggests that large-scale speech pretraining captures reusable structure
in acoustic signals rather than speech-specific information alone, raising the question:
\textit{how far does this transfer extend?}

Animal vocalizations provide a particularly strong test of this question.
Human speech and animal vocalizations are different communication systems,
but both are acoustic signals shaped by temporal structure, spectral content,
and source-specific characteristics. It is therefore natural to ask whether
representations learned from human speech remain useful across species.
This is relevant in bioacoustics, where labeled datasets
are often small, particularly for rare species and individual-level tasks~\cite{knight2024individual}.

Prior work has established that speech representations can transfer to animal
vocalizations. Sarkar and Magimai-Doss~\cite{sarkar2023animal,sarkar2025comparing}
showed that speech-pretrained SSL models encode animal caller information and
later found only marginal benefits from animal-domain pretraining across several
bioacoustic tasks. Cauzinille et al.~\cite{cauzinille2025crossing} extended this
analysis across taxa and tasks, revealing strong layer dependence and suggesting
multilinguality and acoustic robustness as possible factors in transfer.
Complementarily, Miron et al.~\cite{miron2026avex} showed that architecture,
training strategy, and data composition strongly affect bioacoustic representation
transferability. Together, these studies leave open \textit{which properties of
human-speech representations and their pretraining determine successful
cross-species transfer}.

Some factors that may influence such transferability are: language coverage
may encourage representations less specialized to the phonetics of a single language;
animal-domain pretraining may better match the target acoustics; and representations
optimized for human speaker verification may preserve general patterns
relevant to animal identity.
In this work, we examine these factors by comparing foundation models spanning monolingual and multilingual speech, 
speaker verification, animal bioacoustics, and general audio across four species: rock 
hyrax (\textit{Procavia capensis}), zebra finch (\textit{Taeniopygia guttata}), little owl (\textit{Athene noctua}), and infant marmoset (\textit{Callithrix jacchus}). 
Our main contributions are: 

(1) Benchmarking 15 frozen encoders on two tasks: caller identification 
and call-type classification.

(2) Testing the effects of language coverage, pretraining domain, and speaker specialization through controlled model comparisons. 

(3) Probing representations layer by layer to characterize where transferable animal-relevant information emerges.
\begin{figure*}[!t]
\centering
\includegraphics[width=0.85\textwidth]{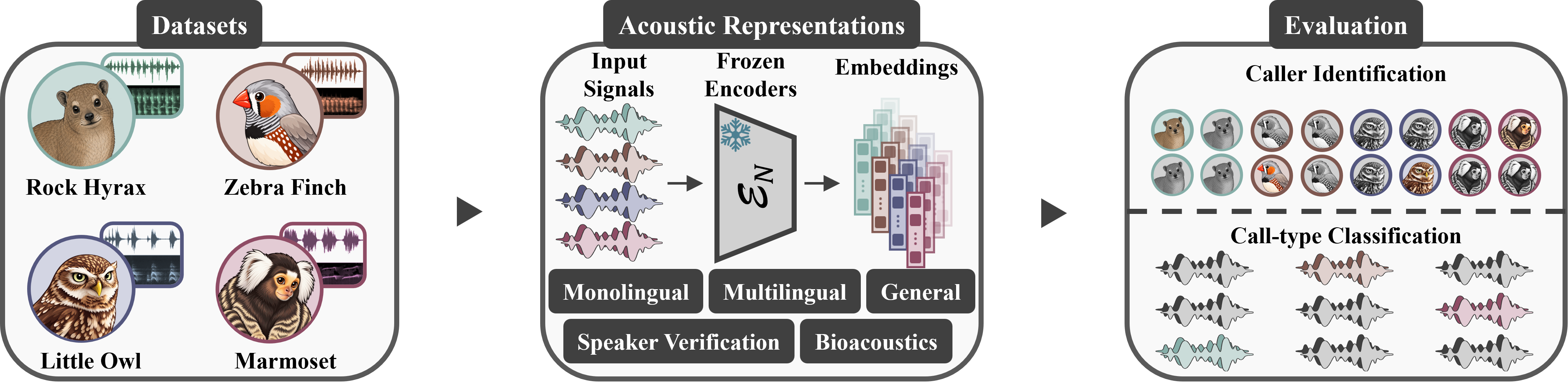}
\caption{Evaluation pipeline. We compare pre-trained encoders spanning monolingual (English) and multilingual speech, speaker verification, animal bioacoustics, and general audio across two tasks: caller identification and call-type classification.}
\label{fig:pipeline}
\end{figure*}

\section{Materials \& Methods}
\label{sec:method}
Figure~\ref{fig:pipeline} summarizes the evaluation pipeline.
Caller identification is evaluated on four datasets and call-type
classification on three, since the little owl corpus contains only one
call type in the subset considered here.
\subsection{Datasets}
\label{ssec:data}
Table~\ref{tab:corpora} summarizes the four datasets.

\textit{Rock hyrax}. Recordings originate from a long-term study at the Ein Gedi
Nature Reserve, Israel, where 18 male hyraxes were recorded during 2002--2013~\cite{demartsev2023rhythm}. The study population was monitored at two
sites within the reserve~\cite{koren2009complex}, and animals were individually
marked and identified during field observations.
Hyrax vocalizations are organized into bouts composed of three
elements: \emph{wails}, \emph{chucks}, and \emph{snorts}.

\textit{Zebra finch}. Recordings come from a captive colony at UC Berkeley and
were collected between 2011 and 2013~\cite{elie2018}. In total, 45 birds were recorded in a
soundproof booth across 147 sessions. Caller identity was labeled during recording,
and calls overlapping with another bird were excluded. 

\textit{Little owl}. Recordings consist of single territorial calls collected from 16 males
in farmland in northern Bohemia and eastern Hungary during 2013--2014~\cite{stowell2018}. 
Birds were recorded with a shotgun microphone and solid-state 
recorder at distances of up to 50\,m, following a one-minute playback. 

\textit{Infant marmoset}. Recordings come from the InfantMarmosetsVox corpus,
comprising five pairs of twins recorded individually in two soundproof rooms
~\cite{zhang2018marmoset}. 
The source corpus contains approximately 58\,h of annotated vocalization onset, offset, call type, and caller identity. 


\begin{table}[!h]
\centering
\caption{Datasets considered in this study.}
\label{tab:corpora}
\setlength{\tabcolsep}{3pt}
\small
\resizebox{0.9\linewidth}{!}{
\begin{tabular}{lcccc}
\hline\hline
& \textbf{Hyrax} & \textbf{Finch} & \textbf{Owl} & \textbf{Marmoset} \\
\hline
Individuals$^\dagger$ & 16 & 33 & 16 & 10 \\
Caller-ID unit & bout & call & call & call \\
Call-type classes$^\ast$ & 3 & 10 & 1 & 10 \\
Caller-ID units & 3{,}863 & 3{,}090 & 952 & 72{,}921 \\
Call-type units$^\dagger$ & 19{,}970 & 3{,}071 & 952 & 72{,}898 \\
Evaluated duration & 89\,min & 21\,min & 20\,min & 464\,min \\
\hline\hline
\end{tabular}}
\vspace{0.5mm}

{\footnotesize \raggedright 
$^\ast$For rock hyrax, the call-type refers to vocal elements (wail, chuck, snort).\par
$^\dagger$Section~\ref{ssec:folds} describes the reasons for dropping some individuals and calls.
}
\end{table}

\textbf{Pre-processing.} 
We standardize recordings to 16\,kHz mono and segment them into vocal units using their respective manual annotations. 
%
Segments shorter than the 0.4\,s encoder input minimum are extended with surrounding audio for hyrax bouts and marmoset calls, and zero-padded otherwise.
%
We also evaluated a bioacoustic denoising model to reduce differences
in recording conditions. However, qualitative inspection suggested that identity-relevant 
frequency regions were being suppressed and was therefore not used.

\subsection{Acoustic representations}

We evaluate English speech models (wav2vec\,2.0 large-lv60~\cite{baevski2020wav2vec},
HuBERT-base~\cite{hsu2021hubert}, WavLM-base+~\cite{chen2022wavlm}, and Whisper-small.en~\cite{radford2023robust}), multilingual counterparts (XLS-R~\cite{babu2022xlsr}, mHuBERT~\cite{boito2024mhubert}, and XEUS~\cite{chen2024xeus}), speaker-verification models (wav2vec2-SV, UniSpeech-SAT-SV~\cite{chen2022unispeech}, WavLM-SV, ECAPA-TDNN~\cite{desplanques2020ecapa}), animal-pretrained models (AVES-bio~\cite{aves}, AVES2-all, and AVES2-bio~\cite{miron2026avex}), and BEATs~\cite{beats} as a general-audio model.
The monolingual models wav2vec 2.0, HuBERT, and WavLM are SSL with a raw-waveform front end, while Whisper is trained for ASR with a log-mel spectrogram (+1D convolutions) front end.
XLS-R-300m and mHuBERT-147 are architecture-matched to wav2vec\,2.0 and
HuBERT, respectively.
For speaker verification, all models except ECAPA-TDNN combine an SSL encoder with an x-vector head; ECAPA-TDNN provides a single speaker embedding. WavLM-SV's encoder matches WavLM-base+.
AVES-bio has a HuBERT-based encoder pretrained on bioacoustic audio, while the BEATs/AVES2-all/AVES2-bio series enables a controlled comparison of pretraining domain with the backbone fixed.
All embeddings are pooled by mean and standard deviation, resulting in a 2048-D feature vector for XEUS, XLS-R, and wav2vec\,2.0 large, 384-D for ECAPA, and 1538-D for the rest.

\textbf{Baseline.}
As a non-pre-trained acoustic baseline, we use 20 Mel-frequency cepstral coefficients (MFCCs) with first- and second-order derivatives from 25\,ms windows
every 10\,ms, mean-and-standard-deviation pooled and concatenated with
log-duration, resulting in a 121-D feature vector. 

\subsection{Evaluation}
\label{ssec:folds}
We perform classification using 3-fold cross-validation.

\textbf{Data splits.}
For caller identification, folds are grouped by recording session.
Individuals with $<3$ sessions have their largest session split into contiguous
blocks so that each individual appears in all folds. This affects 2/16 hyraxes,
13/33 zebra finches, and all 16 little owls, whose corpus has only two
recording partitions; no marmoset is affected
(Section~\ref{sec:results}).
For call-type classification, folds are grouped by caller, so evaluation is
performed on unseen individuals. In both tasks, every target class must contain
at least five samples per fold to avoid unstable per-class recall estimates.

\textbf{Classification.}
Embeddings are standardized and classified with a logistic regression using 
balanced class weights. Layer and $C$ are selected jointly by 2-fold cross-validation 
within each training fold, then refit on the full training split and evaluated 
once on the held-out fold. Performance is reported as mean UAR across folds.
\section{Experiments \& Results}
\label{sec:results}

Tables~\ref{tab:zeroshot} and~\ref{tab:calltype} summarize the
results for caller identification and call-type classification,
respectively. 
For each encoder, we report the UAR at the layer selected on the inner
folds of each split (Section~\ref{ssec:folds}).
MFCCs provide a reference under the same evaluation protocol. 
Pairwise differences are assessed by a paired cluster bootstrap that
resamples recording sessions (caller identification) or callers (call-type
classification) within each fold. All $p$-values reported below are
Holm--Bonferroni adjusted within each comparison family.

\begin{table}[h]
\centering
\caption{Caller identification UARs; bold marks the best encoder per corpus.}
\label{tab:zeroshot}
\setlength{\tabcolsep}{3pt}
\small
\resizebox{0.95\linewidth}{!}{
\begin{tabular}{llcccc}
\hline\hline
\textbf{Domain} & \textbf{Acoustic} & \textbf{Hyrax} & \textbf{Finch} & \textbf{Owl} & \textbf{Marmoset} \\
 & \textbf{representation} & \textbf{n=16} & \textbf{n=33} & \textbf{n=16} & \textbf{n=10} \\
\hline\hline
Baseline & MFCC & 0.328 & 0.455 & 0.649 & 0.261 \\
\hline
Monolingual & HuBERT-base & 0.373 & 0.480 & 0.710 & 0.403 \\
& wav2vec2-large & 0.408 & 0.530 & 0.725 & 0.421 \\
& WavLM-base+ & 0.392 & 0.497 & 0.716 & 0.391 \\
& Whisper-small.en & 0.456 & \textbf{0.581} & 0.708 & \textbf{0.482} \\
\hline
Multilingual & mHuBERT-147 & 0.384 & 0.521 & 0.700 & 0.389 \\
& XEUS & 0.342 & 0.205 & 0.656 & 0.326 \\
& XLS-R-300m & 0.400 & 0.536 & 0.729 & 0.414 \\
\hline
Speaker& ECAPA-TDNN & 0.253 & 0.375 & 0.508 & 0.260 \\
verification& UniSpeech & 0.372 & 0.464 & 0.695 & 0.412 \\
& wav2vec2-SV & 0.374 & 0.443 & 0.681 & 0.339 \\
& WavLM-SV & 0.392 & 0.497 & 0.716 & 0.391 \\
\hline
Animal & AVES-bio & 0.421 & 0.485 & 0.725 & 0.389 \\
& AVES2-all & 0.445 & 0.459 & 0.734 & 0.409 \\
& AVES2-bio & \textbf{0.474} & 0.468 & 0.745 & 0.415 \\
\hline
General & BEATs & 0.456 & 0.469 & \textbf{0.757} & 0.433 \\
\hline\hline
\end{tabular}}
\vspace{0.5mm}
\end{table}

\begin{table}[h]
\centering
\caption{Call-type classification UARs; bold marks the best encoder per corpus.}
\label{tab:calltype}
\setlength{\tabcolsep}{3pt}
\small
\resizebox{0.85\linewidth}{!}{
\begin{tabular}{llccc}
\hline\hline
\textbf{Domain} & \textbf{Acoustic} & \textbf{Hyrax} & \textbf{Finch}  & \textbf{Marmoset} \\
 & \textbf{representation} & \textbf{n=3} & \textbf{n=10} & \textbf{n=10} \\
\hline\hline
Baseline & MFCC & 0.873 & 0.780 & 0.579 \\
\hline
Monolingual & HuBERT-base & 0.911 & 0.822 & 0.699 \\
& wav2vec2-large & 0.929 & 0.828 & 0.664 \\
& WavLM-base+ & 0.931 & 0.799 & 0.689 \\
& Whisper-small.en & 0.913 & \textbf{0.866} & 0.717 \\
\hline
Multilingual & mHuBERT-147 & 0.932 & 0.827 & 0.698 \\
& XEUS & 0.635 & 0.501 & 0.515 \\
& XLS-R-300m & \textbf{0.939} & 0.827 & 0.654 \\
\hline
Speaker & ECAPA-TDNN & 0.865 & 0.729 & 0.469 \\
verification & UniSpeech & 0.918 & 0.810 & 0.708 \\
& wav2vec2-SV & 0.908 & 0.818 & 0.655 \\
& WavLM-SV & 0.931 & 0.799 & 0.689 \\
\hline
Animal & AVES-bio & 0.923 & 0.802 & 0.681 \\
& AVES2-all & 0.860 & 0.771 & \textbf{0.731} \\
& AVES2-bio & 0.858 & 0.795 & 0.726 \\
\hline
General & BEATs & 0.882 & 0.802 & 0.717 \\
\hline\hline
\end{tabular}}
\vspace{0.5mm}

\end{table}

\begin{figure*}[!t]
\centering
\includegraphics[width=\linewidth]{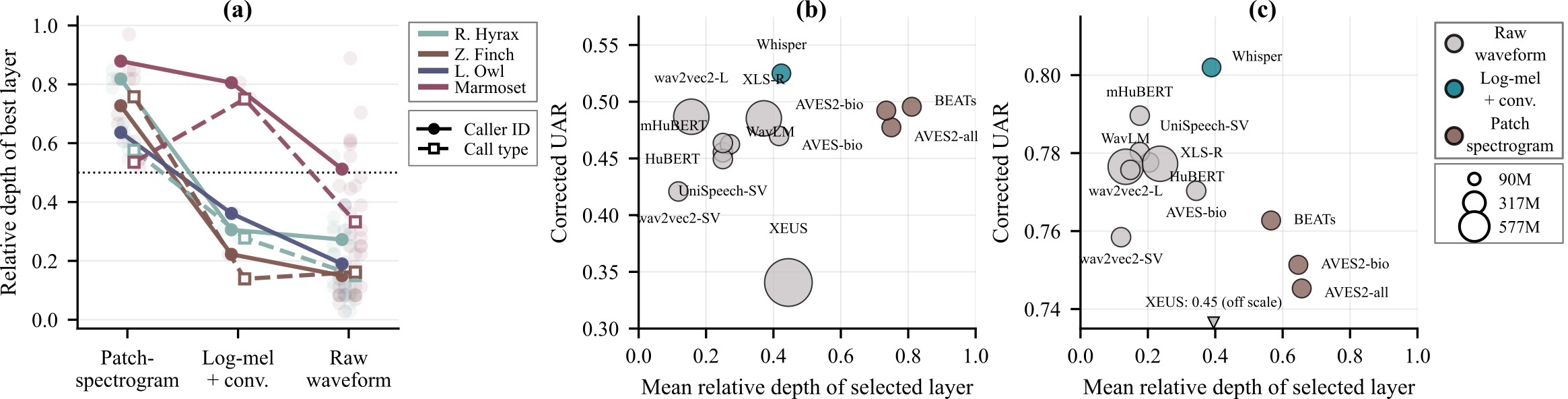}
\caption{(a) Mean selected relative layer depth by front end and task. (b,c) Mean selected depth versus chance-corrected UAR for caller ID and call-type classification. Bubble area denotes model size; color denotes front end. XEUS is off-scale in (c).
}
\label{fig:layers}
\end{figure*}

\textbf{Cross-species transfer.}
We first compare the best speech-pretrained and animal-pretrained
representations in each task--dataset setting, both selected on the inner
folds.
Speech models are significantly better in three of the seven settings: on
zebra finch for caller identification ($+0.119$ UAR, $p<0.001$) and
call-type classification ($+0.036$, $p=0.003$), and on marmoset for caller
identification ($+0.086$, $p<0.001$). Whisper is the selected speech encoder
in all three settings, in every fold but one.
The remaining four settings show no significant difference: hyrax caller
identification ($-0.019$, $p=0.17$), little-owl caller identification
($-0.027$, $p=0.21$), hyrax call-type classification ($+0.006$, $p=0.20$),
and marmoset call-type classification ($-0.014$, $p=0.07$).
In no setting is the animal encoder significantly better.


\textbf{Language coverage.}
Across the 14 architecture-matched HuBERT--mHuBERT and wav2vec\,2.0--XLS-R
comparisons, absolute differences never exceed $0.040$ UAR.
Three are significant, all favoring the multilingual model: mHuBERT on
zebra finch caller identification ($+0.040$, $p<0.001$) and hyrax call-type
classification ($+0.021$, $p=0.002$), and XLS-R on hyrax call-type
classification ($+0.010$, $p<0.001$). The remaining eleven, including all
four marmoset and both little-owl comparisons, are not significant.
Thus, increased language coverage provides at most a small and inconsistent
advantage for cross-species transfer.

\textbf{Pretraining domain.}
With the BEATs backbone fixed, differences among BEATs, AVES2-all, and
AVES2-bio are at most $0.029$ UAR for caller identification and $0.031$ for
call-type classification. Of 14 comparisons against BEATs, five are
significant: four favor BEATs (hyrax call type and
marmoset caller identification, both variants, $p\leq0.017$) and one favors
an animal-pretrained variant (AVES2-all on marmoset call type, $+0.014$,
$p<0.001$).
Animal-domain pretraining therefore provides no consistent advantage.

\textbf{Speaker specialization.}
Speaker-verification models transfer like other raw-waveform speech models. 
WavLM-SV's encoder matches WavLM-base+. Their x-vector heads, in contrast, 
score below every encoder layer of the same model in all 21 settings, by $0.10$--$0.32$
UAR ($p<0.001$ in all cases), and below the MFCC baseline in 26 of 28 settings.
ECAPA-TDNN, which provides only one embedding, is below MFCC in all seven.
Speaker-verification layers therefore transfer poorly, while the
layers beneath them do not; however, adapting the encoder could yield different behavior.

\begin{figure}[h]
\centering
\includegraphics[width=\linewidth]{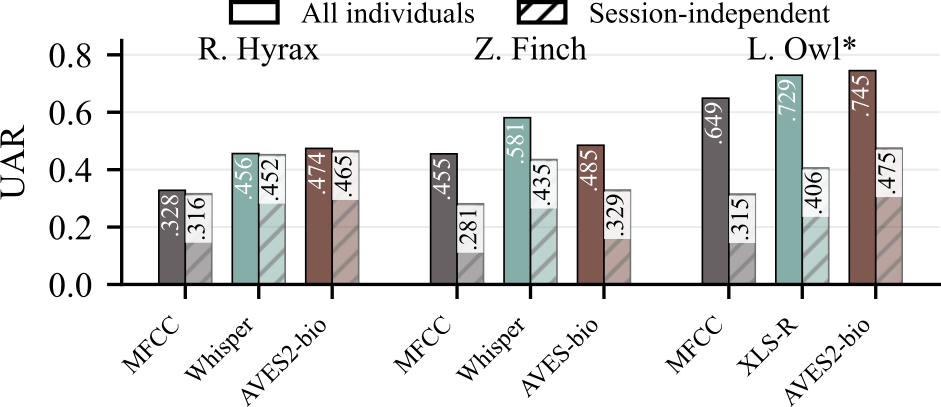}
\caption{Caller identification UAR for all individuals (solid) and for
session-independent individuals only.
$^\ast$Little owl: two-fold across its two recording partitions.}
\label{fig:session}
\end{figure}

\textbf{Session independence.}
Figure~\ref{fig:session} shows the results when the caller 
identification experiments are restricted to session-independent individuals.
Hyrax is essentially unchanged. For zebra finch, the corpus UAR
of $0.581$ decomposes into $0.81$ for the 13 block-split birds and $0.44$ for
the 20 session-independent ones. Little owl drops by $0.19$--$0.37$ for every
encoder. 
Call-type folds hold out callers and are unaffected.

\textbf{Layer-wise transfer.}
A pattern we found across encoders is that selected layer depth is more strongly
associated with the front-end representation than with the pretraining domain
(Figure~\ref{fig:layers}a). For caller identification, raw-waveform models
(e.g. wav2vec) select earlier layers ($0.15$--$0.51$ of relative network depth,
averaged across folds), whereas patch-spectrogram BEATs/AVES2 models select
deeper ones ($0.64$--$0.88$). Animal-pretrained AVES-bio follows the
raw-waveform speech models, while BEATs and AVES2, which share the same front
end, select similar depths. Whisper falls between these groups on caller
identification and varies across datasets. Call-type classification shows the
same pattern on hyrax and zebra finch, but less clearly on marmoset.

Figures~\ref{fig:layers}b and~\ref{fig:layers}c show each encoder by its mean
depth and mean chance-corrected UAR i.e., (UAR $-$ chance)/(1 $-$ chance). 
Whisper is the strongest encoder in both
tasks at intermediate depth. For call-type classification, all raw-waveform
speech encoders cluster in the first quarter of the network within $0.03$
corrected UAR of each other, while the patch-spectrogram encoders sit deeper
and slightly lower. Model size shows no relation to transfer: XEUS is the
largest encoder and the weakest in both tasks.
Since the front end is confounded with model family, this should be interpreted
as an association rather than a controlled comparison. Nevertheless, the
results show that evaluating only the final layer can mischaracterize
cross-species transfer.


\section{Conclusion}

Human-speech representations transfer across species for both caller identity
and call type: the strongest speech models significantly outperform
animal-pretrained models in three of seven settings and do not differ
significantly in the remaining four. For hyrax and marmoset, this holds across
recording sessions, whereas zebra finch and little-owl caller identification
partly reflects within-session recognition.

Controlled comparisons show no systematic advantage from increased language
coverage or animal-domain pretraining. Speaker-verification embeddings
transfer poorly to animal identity, although useful information remains in
their underlying encoders. Transfer is also strongly layer-dependent, with the
most informative representation occurring at different depths across front
ends.

Because all encoders are evaluated with frozen weights, these results
characterize information directly accessible from pretrained representations,
not the full potential of their architectures. Supervised adaptation may therefore 
change these conclusions, as prior work shows that post-training SSL 
encoders can substantially improve bioacoustic transfer~\cite{miron2026avex}.

\textbf{Acknowledgements:}
This work was supported by Deutsche Forschungsgemeinschaft (DFG) under grant number AR 1758/1-1. The authors also thank Dr. Demartsev for curating the rock hyrax data.

\textbf{Compliance with Ethical Standards:}
We performed no new animal procedures for this work.
All recordings used were collected in earlier studies under the ethical approvals reported therein (see section~\ref{ssec:data}) and were obtained for the present analysis under the terms under which their authors made them available. 

\bibliographystyle{IEEEbib}
\bibliography{refs}

\end{document}